\documentstyle[12pt,aps,epsfig]{revtex}
\newcommand{\be}{\begin{equation}}
\newcommand{\ee}{\end{equation}}
\newcommand{\bea}{\begin{eqnarray}}
\newcommand{\eea}{\end{eqnarray}}

\def\Journal#1#2#3#4{{#1} {\bf #2}, #3 (#4)}

\def\NP{{Nucl. Phys.}}
\def\PL{{Phys. Lett.} B}
\def\PR{{Phys. Rep.}}
\def\PRL{Phys. Rev. Lett.}
\def\PRD{{Phys. Rev.} D}
\def\PRC{{Phys. Rev.} C}

\def\ZPC{{Z. Phys.} C}

\def\AP{{Ann. Phys. (N.Y.)}}

\begin{document}
\title{Effect of quark-jet energy loss on direct photons
in ultrarelativistic heavy-ion collisions\footnote[1]{Supported by
BMBF, DFG, GSI, and DAAD.}}
\author{A. Dumitru}
\address{Physics Department, Yale University\\
P.O.\ Box 208124, New Haven, CT 06520, USA\\}
\author{N. Hammon}
\address{Institut f\"ur Theoretische Physik, J.W. Goethe Universit\"at\\
Robert-Mayer Str. 8-10, 60054 Frankfurt, Germany}
\date{\today}
\maketitle   
\begin{abstract}
We discuss the transverse momentum distribution of thermal and prompt
photons in ultrarelativistic heavy ion collisions.
The prompt photons are mostly produced by fragmentation of quark jets
(bremsstrahlung). If these quark jets suffer a large
energy loss in heavy-ion reactions, prompt photons are significantly
suppressed. Thermal electromagnetic radiation from the quark-gluon plasma and
the hadron gas might then dominate the intermediate $k_T$-range.
In central Au+Au collisions at $\sqrt{s}=200A$~GeV, and in the
range $2~{\rm GeV}<k_T<4~{\rm GeV}$, the inverse slope of the thermal
radiation is
$360-460$~MeV, while that of the QCD photons is $500-1000$~MeV. This large
difference might allow to disentangle the various sources of direct photons
experimentally. At $\sqrt{s}=5.5A$~TeV, nuclear shadowing and quark-jet
quenching suppresses the QCD-photons strongly for $k_T<5$~GeV. Thermal
radiation dominates in this $k_T$-region.
\end{abstract}
\newpage
In central collisions of heavy nuclei at ultrarelativistic energies
a very hot and dense system of strongly interacting matter can be
produced. By detecting produced particles like real and virtual photons
or $J/\Psi$ mesons one hopes to probe this short-lived hot and dense state
(see e.g.~\cite{JHBM}).

Electromagnetic radiation is of particular interest since (once produced)
it leaves the hot and dense region without further
interactions~\cite{McT}. Thus, it can provide information about the state of
the source (e.g.\ its temperature or expansion velocity, in case of thermal
radiation) at the space-time point of emission.

In particular, it has been suggested~\cite{Shuryak,Ruusk,Sriv}
that thermal photons might be detectable in the transverse momentum window
$k_T=2-5$~GeV, provided that photons from final-state decays of light mesons
can be reliably identified and subtracted. However,
even at the BNL-RHIC energy, $\sqrt{s}=200A$~GeV, prompt photons
(produced initially in reactions between the partons of the incoming
nuclei) might dominate over the thermal radiation down to $k_T=2$~GeV
\cite{DirPhot,Hwa}. Only at the CERN-LHC energy,
$\sqrt{s}=5.5A$~TeV, might thermal radiation be dominant for $k_T=2-5$~GeV,
which is due to higher initial temperature of the plasma, and shadowing of the
nuclear parton distributions.

If quark-jet quenching is absent in ultrarelativistic
heavy-ion collisions, fragmentation of quark jets into a collinear photon and
a quark (i.e.\ Bremsstrahlung) contributes significantly to the transverse
momentum spectrum of prompt photons at midrapidity~\cite{DirPhot,Hwa},
in addition to the QCD-Compton and quark-antiquark annihilation processes.

On the other hand, in the high-multiplicity environment expected at the
future heavy-ion colliders RHIC and LHC, the newly produced partons in the
central region might materialize very rapidly~\cite{parttherm}.
In this case, produced quark
jets could suffer a substantial energy loss while propagating through the
hot and dense central region~\cite{WGP,jquench,KnVo}. This energy loss is
caused by induced gluon radiation. Consequently, their
fragmentation into high-$k_T$ photons would be inhibited.

The prompt photon spectrum should
therefore also be a sensitive probe of energy-loss effects. Moreover, it is
not sensitive to gluon jets and can thus be used to study in particular
quenching of quark-jets.
To give a first estimate of the maximal effect that can be expected,
we compare the prompt photon spectrum with and without the contribution
from quark-jet Bremsstrahlung. The latter corresponds to
the case of fully quenched quark jets, i.e.\ complete energy loss. 
More detailed investigations of the bremsstrahlung spectrum as a function
of the quark-jet energy loss will be presented elsewhere.

The energy loss of high-$p_T$ jets propagating through hot and dense
QCD-matter has been studied in refs.~\cite{WGP,jquench}, cf.\ also
ref.~\cite{KnVo}. {\small HIJING}
simulations have shown~\cite{Wang} that this energy loss might reflect
in a strong suppression of high-$p_T$ pions in heavy-ion
collisions as compared to proton-proton collisions (at the same center-of-mass
energy per nucleon). In ref.~\cite{tagphot} it was proposed to study
jet quenching by measuring the energy of the jet fragments in the opposite
direction of a tagged photon. This assumes that the high-$k_T$
photons are predominantly produced via the Compton process $gq\rightarrow
\gamma q$. On the other hand, if Bremsstrahlung gives an important
contribution, the energy of the photon obviously is not equal to that of
the quark-jet, and thus does not allow for a direct determination of
its energy loss. This study this is restricted to photon transverse momenta
well above $5$~GeV.
The detection of this jet quenching effect is one of
the issues addressed by the STAR experiment~\cite{STAR} at the Relativistic
Heavy-Ion Collider (RHIC), and the single photons will be measured by the
PHENIX experiment~\cite{PHENIX}.

In this letter we do not discuss the mechanism for the energy loss itself
but propose an alternative observable for the quenching of quark jets,
namely prompt photons. The interaction of high-$k_T$ photons with
the hot and dense QCD medium can be neglected~\cite{KLS,Thoma}, they thus
do not suffer any energy loss while propagating through the medium.

To calculate prompt photon production in $p+p$ reactions, we convolute the
cross-section for the given elementary process with the appropriate
parton distribution of the proton, cf.\ e.g.\ refs.~\cite{Hwa,ppGamma}.
We work in LO (with a $K$-factor of
$K=2$ at $\sqrt{s}=200A$~GeV and $K=1.5$ at $\sqrt{s}=5.5A$~TeV to account for
the contribution of higher orders in $\alpha_s$)
and at the twist-2 level, and employ the GRV-95 parton distribution function
parametrizations for the proton~\cite{GRV}. To obtain the photon spectrum
in heavy ion reactions, we multiply the cross section for $p+p$ reactions
with the nuclear overlap functions $T_{AuAu}(b=0)=29{\rm mb}^{-1}$ and
$T_{PbPb}(b=0)=32{\rm mb}^{-1}$, respectively. We account for nuclear
shadowing effects by multiplying with $R_{F_2}$ as parametrized in
ref.~\cite{Eskola}. The $Q^2$ dependence of $R_{F_2}$ is neglected.

We also compare these spectra to the photons
from the thermalized stage of the reaction. These are computed assuming
longitudinally boost-invariant and cylindrically symmetric transverse
hydrodynamical expansion of the thermalized quark-gluon plasma (QGP). To obtain
an upper estimate for the thermal photons, we assume a very short
thermalization time and a
high initial temperature ($\tau_i=0.124$~fm/c, $T_i=530$~MeV for
Au+Au at RHIC, $\sqrt{s}=200A$~GeV, and $\tau_i=0.1$~fm/c, $T_i=880$~MeV for
Pb+Pb at LHC, $\sqrt{s}=5.5A$~TeV).
For the QGP we assume the MIT bag-model equation of state for two massless
quark flavours and a bag-constant of $B=380$~MeV/fm$^3$. Thus, the total
entropy at midrapidity is $dS/d\eta=5000$ for Au+Au at RHIC, and
$dS/d\eta=18700$ for Pb+Pb at LHC. 

The hadronic phase
is modelled as an ideal gas of massive $\pi$, $\eta$, $\rho$, and $\omega$
mesons. The two equations of state are matched by Gibbs conditions of phase
equilibrium thus leading to a first-order phase transition (at $T_C=160$~MeV).
Finally, the thermal photon production rate derived in ref.~\cite{KLS} is
integrated incoherently over that volume of the forward light-cone with
temperature above $100$~MeV. The Bremsstrahlung contribution to the
{\em thermal} photon emission rate~\cite{KnVo,thBrems} is neglected since it
is important only at lower transverse momenta.
For further details of the calculations
of both the prompt as well as the thermal photons please refer to
ref.~\cite{DirPhot}.

A third source of photons are the final-state decays of $\pi^0$ and
$\eta$ mesons. This background contribution, which has been discussed e.g.\ in
refs.~\cite{Ruusk,Sriv}, turns out to exceed the thermal yield in the
transverse momentum range considered here. Thus, as already argued in
ref.~\cite{Sriv}, these decay photons have to be identified and subtracted
in order that the thermal radiation could be observable. At lower energy
(Pb+Pb collisions at $\sqrt{s}=18A$~GeV, total central entropy $dS/dy
\sim3000-3500$) this has been successfully performed by the WA98
collaboration~\cite{WA98}. We assume that this will be possible also
at the higher energies.

\begin{figure}[htp]
\centerline{\hbox{\epsfig{figure=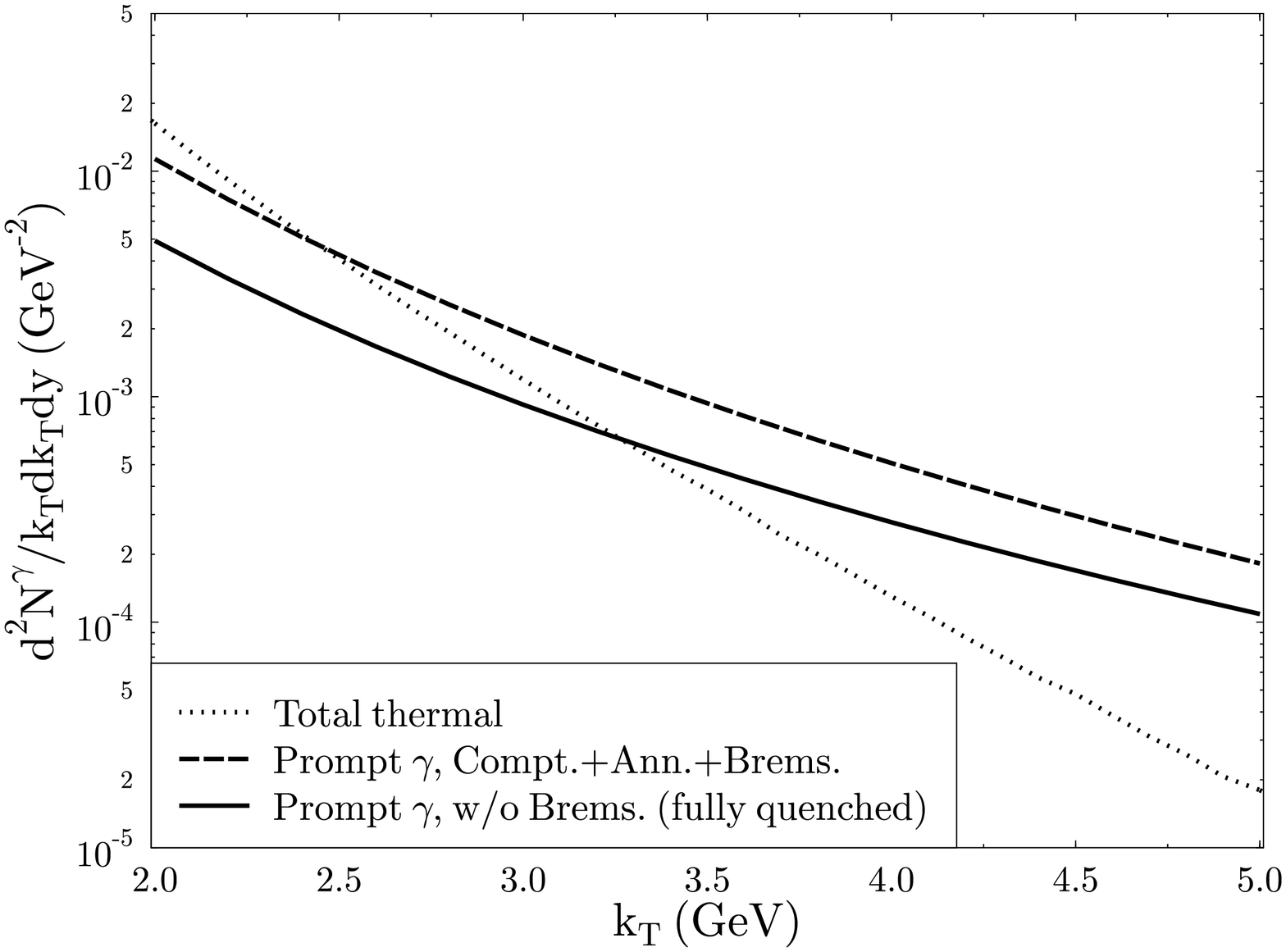,height=9cm,width=12cm}}}
\vspace*{.5cm}
\caption{Transverse momentum distribution of thermal (dotted curve) and prompt
QCD-photons at midrapidity
(solid curve: sum of compton and annihilation contributions;
dashed curve: sum of compton, annihilation, and bremsstrahlung
contributions); for central $Au+Au$ collisions at
RHIC energy.}
\label{photkT}
\end{figure}
Figure~\ref{photkT} shows that at RHIC energy the transverse momentum
spectrum above
$k_T\approx2.5$~GeV is dominated by prompt photons, if quark jet quenching is
absent. Thus, even if decay photons could be identified and
subtracted, the thermal radiation would not be visible (in this range of
$k_T$).

However, in central heavy-ion collisions the prompt photon spectrum might
be modified. While nuclear shadowing effects were found to be small at this
energy~\cite{DirPhot}, quenching of quark jets might suppress
prompt photon production. In the extreme case where the fragmentation of
quark jets into photons is suppressed completely, thermal radiation
shows up again in the transverse momentum range $k_T\le3$~GeV, cf.\
Fig.~\ref{photkT}. For the very short thermalization time and high
initial temperature assumed here, the thermal photon spectrum is dominated
by photons from the QGP. For higher $\tau_i$ and lower $T_i$, however,
the hadronic phase might contribute a similar number of real
photons~\cite{DirPhot}.

At this energy, the thermal and prompt photons have very different
inverse slopes of the transverse momentum distributions. Even for such
small thermalization times as $\tau_i=0.124$~fm, and including
collective transverse expansion, the inverse slope of the thermal photons
is $T^*_{th}\le500$MeV, for $k_T\le5$~GeV. $T^*$ is calculated as
\be
T^* = -\frac{1}{{\rm d/dk_T}\ln \left({\rm d}^2N^\gamma/k_T {\rm d}k_T 
{\rm d}y\right)} \quad.
\ee
More specifically, we obtain $T^*_{th}=360$~MeV at $k_T=2.2$~GeV and
$T^*_{th}=460$~MeV at $k_T=3.8$~GeV. On the other hand, the inverse slopes
of the prompt photons are $T^*_{prompt}=500$~MeV and $T^*_{prompt}=750$~MeV
at $k_T=2.2$~GeV and $k_T=3.8$~GeV,
respectively. This large difference in $T^*$ should make it possible to
distinguish thermal and prompt photons experimentally.

\begin{figure}[htp]
\centerline{\hbox{\epsfig{figure=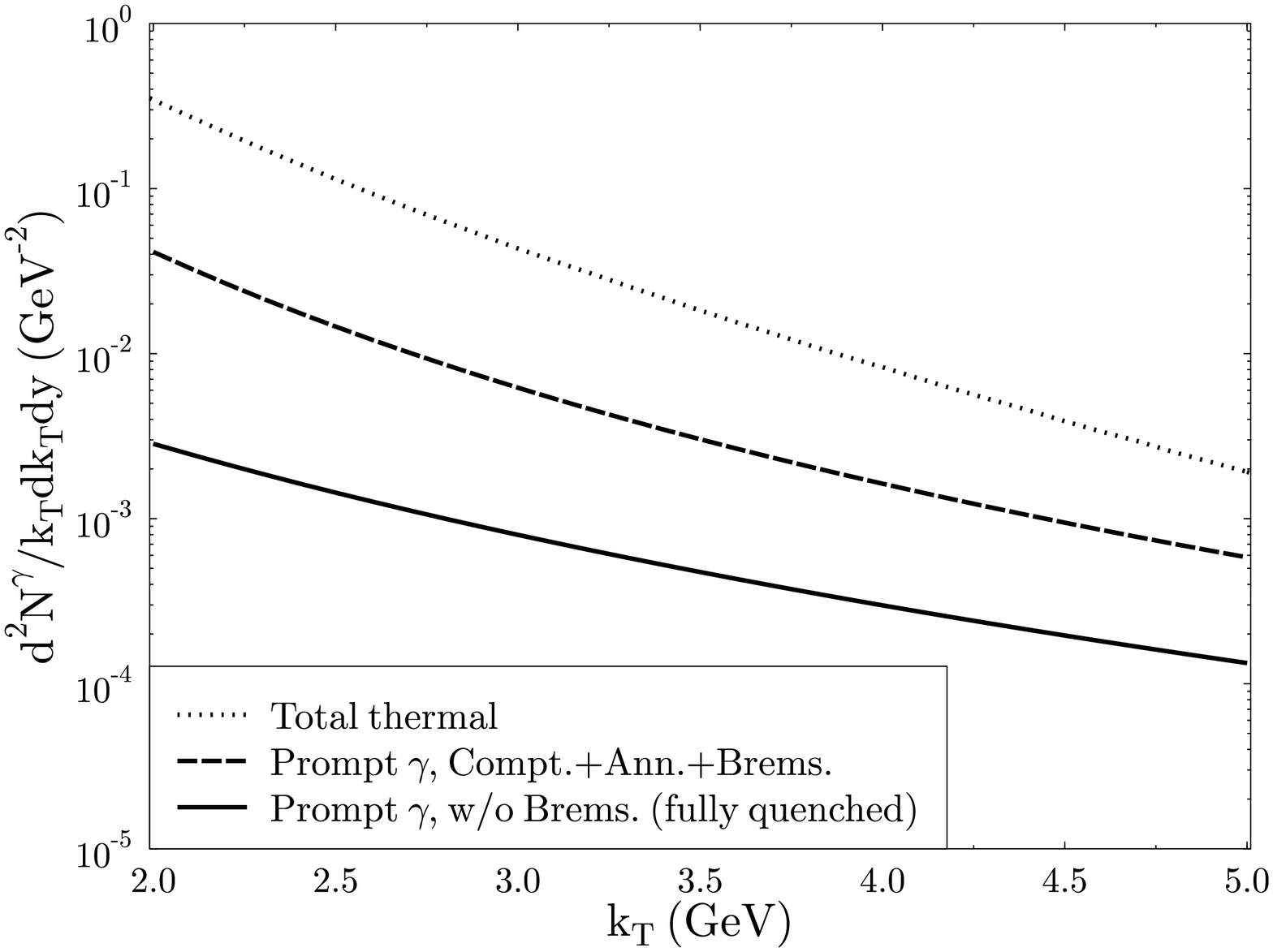,height=9cm,width=12cm}}}
\vspace*{.5cm}
\caption{Transverse momentum distribution of thermal (dotted curve) and prompt
QCD-photons at midrapidity
(solid curve: sum of compton and annihilation contributions;
dashed curve: sum of compton, annihilation, and bremsstrahlung
contributions); for central $Pb+Pb$ collisions at LHC energy.}
\label{photkTlhc}
\end{figure}
Due to shadowing of the nuclear structure functions,
the prompt photons might be below the thermal radiation at LHC energy,
cf.\ Fig.~\ref{photkTlhc}. The energy loss of quark jets reduces the
non-thermal radiation even further. At $k_T=2$~GeV, e.g., the prompt
radiation is two orders of magnitude below the thermal contribution.

In summary, we have pointed out that the spectrum of prompt photons emitted
in ultrarelativistic heavy-ion collisions might be sensitive to
the quark-jet energy loss in the medium.
At RHIC, very strong jet quenching might suppress prompt
photons to below the thermal radiation for not too high transverse
momenta, $k_T\le3$~GeV. The photons with higher transverse momenta
offer the opportunity to study quark-jet quenching in the hot and dense
QCD medium. This is an independent observable of the jet-quenching
effect (in addition to the proposed study of hadron spectra at high
$p_T$~\cite{Wang}).

At the LHC energy, the conditions for detecting
the thermal electromagnetic radiation (from the QGP) are even more
favorable, since high initial temperatures enhance thermal emission, while
nuclear shadowing and quark-jet quenching suppress prompt photon production.

The assumption that the partons formed at midrapidity essentially
thermalize immediately is, of course, rather crude. In principle,
one would expect that a pre-equilibrium stage exists, which would then also
emit photons~\cite{Kaempf}, thus ``interpolating'' between prompt and
thermal radiation. This can be studied, e.g., within the parton cascade
approach~\cite{PCMph}. According to our results, however, coherence
effects~\cite{jquench} on photon production
(up to a few GeV of transverse momentum)
in ultrarelativistic heavy-ion collisions are important and should be
taken into account.
\acknowledgements
We are indebted to M.\ Gyulassy, K.\ Redlich, D.H.\ Rischke, and X.-N.\ Wang
for helpfull discussions.
A.D.\ gratefully acknowledges a postdoctoral fellowship by the German
Academic Exchange Service (DAAD).

\end{document}